\documentclass{desyproc}

\begin{document}
\title{WISPers from the Dark Side: Radio Probes of Axions and Hidden Photons}

\author{{\slshape Dieter Horns$^1$, Axel Lindner$^2$, Andrei Lobanov$^{3,1}$, Andreas Ringwald$^2$}\\[1ex]
$^1$Institut f\"ur Experimentalphysik, Universit\"at Hamburg, Germany\\
$^2$Deutsches Elektronen-Synchrotron (DESY), Hamburg, Germany\\
$^3$Max-Planck-Institut f\"ur Radioastronomie, Bonn, Germany}

\contribID{familyname\_firstname}

\desyproc{DESY-PROC-2013-XX}
\acronym{Patras 2013} 
\doi  

\maketitle

\begin{abstract}
  Measurements in the radio regime embrace a number of effective
  approaches for WISP searches, often covering unique or highly
  complementary ranges of the parameter space compared to those
  explored in other research domains. These measurements
  can be used to search for electromagnetic tracers of the hidden
  photon and axion oscillations, extending down to $\sim 10^{-19}$\,eV
  the range of the hidden photon mass probed, and closing the last
  gaps in the strongly favoured 1--5\,$\mu$eV range for axion dark
  matter. This provides a strong impetus for several new initiatives
  in the field, including the WISP Dark Matter eXperiment (WISPDMX)
  and novel conceptual approaches for broad-band WISP searches in the
  0.1--1000\,$\mu$eV range.
\end{abstract}

\section{WISP in the radio regime}

\begin{figure}[ht!]
\centerline{\includegraphics[width=0.99\textwidth]{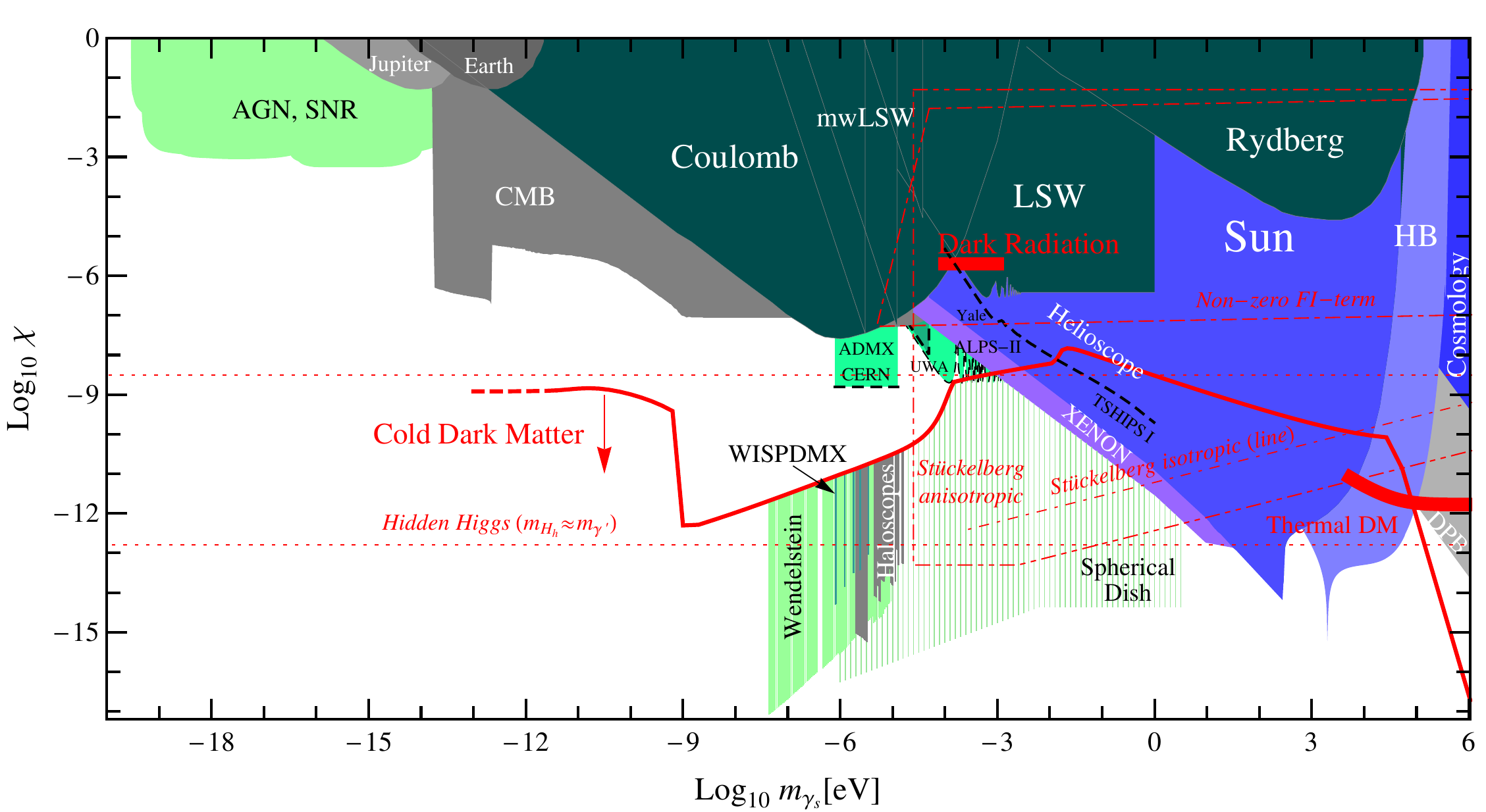}}
\centerline{\includegraphics[width=0.99\textwidth]{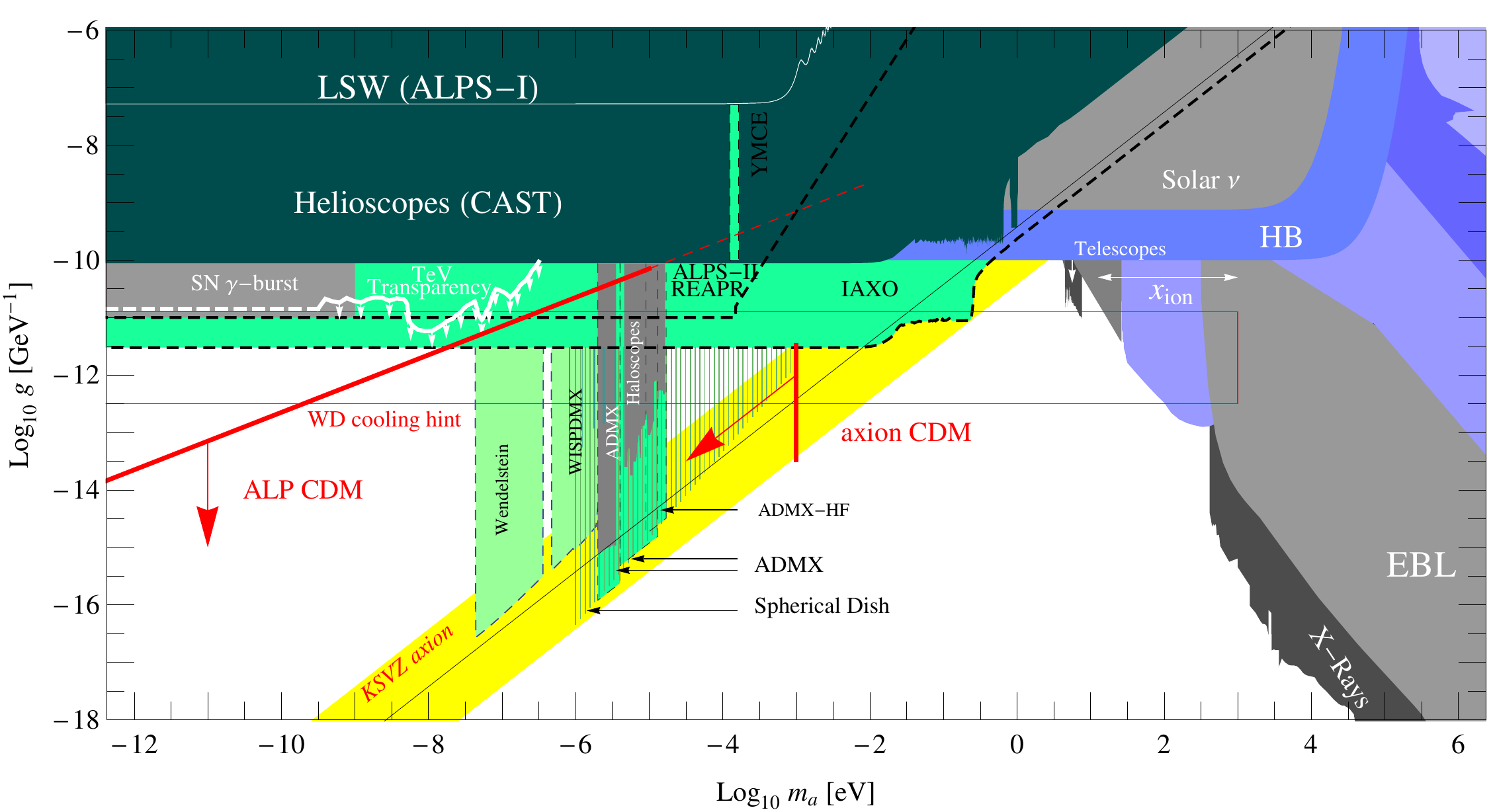}}
\caption{Exclusion limits for hidden photon (top) and ALP (bottom)
  couplings to SM photons. Existing measurements are indicated with
  gray/blue/dark green shades and white captions. Expected limits from future
  measurements are indicated with light green shades and black captions. The yellow band in the axion plot
  marks properties of the QCD axion. Red
  color indicates theoretical constrains for hidden photon and axion production
  and expectations for dark matter and dark radiation (for hidden photons) produced by
  hidden photons (figures adapted
  from~\cite{Hewett:2012if}).}\label{fig:lobanov1}
\end{figure}

The scope of experimental studies of dark matter (DM) has been
expanding steadily towards low energies and to weakly interacting slim
particles (WISP)~\cite{Jaeckel:2010ni,Ringwald:2012hr,Hewett:2012if}
such as axions, axion-like particles (ALP) and hidden photons
(HP). Best revealed by their coupling to standard model (SM) photons,
the WISP may give rise to dark matter for a broad range of the
particle mass and the photon coupling strength~\cite{Arias:2012az} as
indicated by red lines in Fig.~\ref{fig:lobanov1}. At particle masses
above $\sim 10^{-3}$\,eV, the existing constraints effectively rule
out WISP as DM particles, while there are very few measurements
reaching sensible exclusion levels at lower energies. This domain
corresponds to the radio regime at frequencies below 240\,GHz where
highly sensitive measurement techniques are developed for
radioastronomical measurements, with typical detection levels of
$\lesssim 10^{-22}$\,W. Such sensitivity provides excellent
opportunities for laboratory
\cite{Asztalos:2009yp,Wagner:2010mi,Horns:2013} and astrophysical
\cite{Pshirkov:2009} searches for WISP of both cosmological (dark
matter) and astrophysical origin (photon-WISP conversion). The
dependence of the latter signal in HP on the distance to the target
object also offers a unique tool for reaching particle masses down to
$\lesssim 10^{-18}$\,eV~\cite{Lobanov:2013pr} (see
Fig.~\ref{fig:lobanov1}).

\section{Astrophysical measurements}

Astrophysical measurements in the radio can broaden substantially the
range of parameter space probed for ALP and HP. Analysis of the WMAP CMB measurements in the radio domain at frequencies above 22\,GHz has already provided excellent ALP and HP bounds down to masses of $2\times
10^{-14}$\,eV~\cite{Mirizzi:2009nq,Mirizzi:2009iz}. Dedicated radio astronomical measurements at frequencies below 22\,GHz
should extend axion searches to masses
below $10^{-9}$\,eV and probe coupling constants down to
$10^{-14}$\,GeV$^{-1}$~\cite{Chelouche:2009}. Signals from relic DM
axions can be detected in the spectra of isolated neutron
stars~\cite{Pshirkov:2009} and possibly also in pulsars.

\subsection{Hidden photon signals in compact radio sources}

For hidden photons ($\gamma_\mathrm{s}$), radio observations at
frequencies below 22\,GHz offer an excellent (if not unique) tool for
placing bounds on the mixing angle $\chi$ for $m_{\gamma_\mathrm{s}}$
down to $\approx 10^{-18}$\,eV~\cite{Lobanov:2013pr}. Existing data
are sufficiently accurate for detection of kinetic mixing angles $\chi$
down to $\sim 0.01$, yielding presently a weak hint for a
possible oscillatory signal with $\chi \approx 0.02$ in the
2--$5\,\times 10^{-16}$\,eV energy range. As adverse systematic effects
mimicking the signal cannot be presently excluded, this
indication should be verified. Placing better bounds on $\chi$ down to
$\lesssim 10^{-3}$ can be made by using the expanded capabilities of
the next generation radio astronomical
facilities~\cite{Lobanov:2013pr}.

\section{Laboratory experiments}

Laboratory experiments using resonant microwave cavities at
frequencies between 0.5 and 34\,GHz have yielded the best sensitivity
achieved for HP and ALP dark matter at masses below
10$^{-3}$\,eV~\cite{Bradley:2003rv}.
While capable of reaching the fundamental sensitivity levels, these
experiments are slow in scanning over large ranges of
mass. Novel and fast broadband measurement
techniques are critically needed here.

\subsection{Microwave cavity experiments}

Building on the success of the ADMX axion DM
searches~\cite{Asztalos:2009yp,Wagner:2010mi,Bradley:2003rv} covering
the 2-5\,$\mu$eV energy range, a WISP Dark Matter eXperiment has been
initiated at DESY and the University of Hamburg, aiming at covering
the 0.8-2\,$\mu$eV energy range. The experiment utilizes a 208-MHz
resonant cavity used at the DESY HERA accelerator and plans to
make use of the H1 solenoid magnet. The cavity has a volume of 460
liters and a resonant amplification factor $Q=46000$ at the ground
TM$_{010}$ mode. The H1 magnet provides $B=1.15$\,T in a volume of
7.2\,m$^3$. The signal is amplified by a broad-band 0.2--1\,GHz
amplifier with a system temperature of 100\,K. Broad-band digitization
and FFT analysis of the signal are performed using a commercial 12-bit
spectral analyzer, enabling measuring at several resonant modes
simultaneously.

Since the bandwidth of a single measurement is $\propto Q^{-1}$, the
resonant modes of the cavity must be tuned in order to enable scanning
over a sizable range of particle mass. The tuning will be done with a
plunger assembly providing a $\sim 2$\,MHz tuning range at the ground
mode. The expected exclusion limits are shown in
Fig.~\ref{fig:lobanov1}.

\subsection{Experimental concepts for broad-band searches}

The exceptional sensitivity of microwave cavity experiments comes at
the expense of rather low scanning speeds ($\sim 10$\,MHz/year for
WISPDMX), which makes it difficult to implement this kind of
measurements for scanning over large ranges of particle mass. To
overcome this difficulty, new experimental concepts are being
developed that could relax the necessity of using the resonant
enhancement and working in a radiometer mode with an effective $Q=1$.

The measurement bandwidth of radiometry experiments is
limited only by the detector technology, with modern detectors
employed in radio astronomy routinely providing bandwidths in excess
of 1 GHz and spectral resolutions of better than $10^{6}$.

One possibility for a radiometer experiment is to employ a spherical
dish reflector that provides a signal enhancement proportional to the
area of the reflector~\cite{Horns:2013}.  Another possibility is to
use the combination of large chamber volume and strong magnetic field
provided by superconducting TOKAMAKs or stellarators such as the
Wendelstein 7-X stellarator in Greifswald (providing $B=3\,T$ in a
30\,m$^3$ volume).  The exclusion limits expected to be achievable with
the spherical reflector and stellarator experiments are shown in
Fig.~\ref{fig:lobanov1}.

Deriving from the stellarator approach, a large chamber can be
designed specifically for the radiometer searches, with the inner
walls of the chamber covered by fractal antenna elements providing a
broad-band receiving response and also enabling directional
sensitivity to the incoming photons (through high time resolution
enabling phase difference measurements between individual elements).

Further exploration of these approaches should ultimately enable
performing definitive searches for hidden photon and axion/ALP dark matter
in the $10^{-7}$--$10^{-3}$\,eV mass range.

\section*{Acknowledgments}

Andrei Lobanov acknowledges support from
the Collaborative Research Center (Sonderforschungsbereich) SFB 676
``Particles, Strings, and the Early Universe'' funded by the German
Research Society (Deutsche Forschungsgemeinschaft, DFG).


\begin{footnotesize}

\end{footnotesize}


\end{document}